\documentstyle[prb,eqsecnum,aps]{revtex}
\begin{document}

\title{Phase Transitions in One-Dimensional Truncated Bosonic Hubbard Model \\
and Its Spin-1 Analog}

\author{Vladimir A.\ Kashurnikov and Andrei V.\ Krasavin}

\address{Moscow State Engineering Physics Institute, 
115409 Moscow, Russia }

\author{Boris V.\ Svistunov}

\address{
Russian Research Center ``Kurchatov Institute", 123182 Moscow, Russia}

\maketitle
\begin{abstract}
We study one-dimensional truncated (no more than 2 particles on a site)
bosonic Hubbard model in both repulsive and attractive regimes by exact
diagonalization and exact worldline Monte Carlo simulation. In the
commensurate case (one particle per site) we demonstrate that the point of
Mott-insulator -- superfluid transition, $(U/t)_c=0.50\pm 0.05$, is
remarkably far from that of the full model. In the attractive region we
observe the phase transition from one-particle superfluid to two-particle
one. The paring gap demonstrates a linear behavior in the vicinity of the
critical point. The critical state features marginal response to the gauge
phase. We argue that the two-particle superfluid is a macroscopic analog of
a peculiar phase observed earlier in a spin-1 model with axial anisotropy.
\end{abstract}

\vspace{0.5cm}


\section{Introduction}
\label{sec:1}

In the family of Hamiltonians, which the physics of strongly correlated
systems deals with, the one-dimensional bosonic Hubbard one, 
\begin{equation}
H=\sum_{i=1}^{N_a}\left\{ -t(a_i^{\dag }a_{i+1}+\mbox{H.c.})+{\frac U2}%
n_i(n_i-1)\right\} \;,  \label{Hub}
\end{equation}
[$a_i^{\dag }$ creates a particle on the site $i$, $t$ is the hopping
amplitude, $n_i=a_i^{\dag }a_i$, $N_a$ is the number of sites] plays an
important part as a laconic model for the superfluid (SF) -- Mott-insulator
(MI) phase transition \cite{Fisher,Scalettar,Batrouni} (and for SF --
Bose-glass (BG) and BG--MI transitions in the presence of disorder \cite
{Fisher,Giamarchi}). Along with the full model Eq.\ (\ref{Hub}) sometimes
actually its truncated counterpart, defined by the requirement that site
occupation numbers are less than $3$, is considered: either deliberately,
for the purpose of simplification \cite{Singh}, or implicitly, in
approximate treatments which do not distinguish between the full and the
truncated models \cite{Freericks,Krauth}. One ordinarily motivates such a
replacement by the observation that in the full model even at the boundary
point $U=0$ [At $U<0$ the system is unstable against a collapse of almost
all the particles onto one site.] the quantum-mechanical probability to find
out that the occupation number of a given site is greater than $2$ equals to 
$(e-5/2)/e\approx 0.080$, so it is very likely that the triply (and more)
occupied sites play no essential role.

However, a rather accurate method ``exact diagonalization plus
renormalization group (RG)'' \cite{Kashurnikov} demonstrated the absence of
the superfluid phase in the truncated model almost down to $U=0$, in sharp
contrast to the above-mentioned reasoning, and to the results of a number of
approximate treatments \cite{Singh,Freericks,Krauth}. It is essential that
the conclusion of Ref.\ \cite{Kashurnikov} about the absence of
superfluidity in the macroscopic limit is not a direct one. What was really
demonstrated is that at arbitrarily small $U$ the known RG flow of
superfluid parameters, initialized numerically at intermediate scales, leads
to the destruction of superfluidity in favor of Mott insulator. Hence, it
would be instructive to check this fact explicitly by Monte Carlo (MC)
simulation of a sufficiently large system, allowing direct measuring the
insulating gap at $U\ll 1$. This is the primary goal of the present paper.

Another interesting aspect of the truncated bosonic Hubbard model,
distinguishing it from the full one, is that at $U<0$ it is stable against
the collapse onto one site. Therefore our secondary goal is to investigate
the region of attraction: First, one may expect to find here the superfluid
phase. Secondly, at a sufficiently strong attraction a pairing of particles
should take place, as is clear from the perturbative analysis at $%
U\rightarrow -\infty $, and it is interesting to study numerically the
corresponding phase transition. Note that in Ref. \cite{Stoof} the close
problem of the BCS-like pairing was examined in a dilute atomic Bose gas
with an effective attractive interaction.

Finally, 1D truncated bosonic Hubbard model is not of academic interest
only. Macroscopically, it is very close to a spin-$1$ (anisotropic) chain in
an easy-axis ($U<0$), or easy-plane ($U>0$) environment. Its study thus
provides deep insight into the physics of phase transitions in the spin
chains.

In the present paper we report a study of the model by the exact
continuous-time worldline MC approach, developed recently by Prokof'ev,
Svistunov and Tupitsyn \cite{PST1,PST2}. (Some preliminary results were
obtained by the standard worldline MC method. For some special purposes we
also employ exact diagonalization.) In the commensurate case we have found
the critical value of the MI--SF transition, $(U/t)_c=0.50\pm 0.05$, which
is more than $6$ times lower than that of the full model. In the attractive
region, $U<0$, we observe a specific pairing phase transition when the
system passes from the one-particle superfluid vacuum to the two-particle
one. At the filling factor equal to unity [commensurability being not
relevant] the transition occurs at $U/t\approx -6.0$. The critical behavior
is characterized by linear collapse of the pairing gap as a function of $U$,
and by a marginal response to the gauge phase at the transition point.

We also consider spin-1 analog of the truncated bosonic model, which was
studied earlier by exact diagonalization \cite{Botet,Solyom}. The question
of our interest here is a peculiar phase which was observed in Ref.\ \cite
{Botet}, but was not revealed in Ref.\ \cite{Solyom}, and, as far as we
know, was never discussed since then. We demonstrate that the phase does
really exist, being a direct macroscopic analog of the two-particle
superfluid phase of the bosonic system.


\section{Repulsive Regime}

\label{sec:2}

The MI--SF transition in the truncated one-dimensional commensurate bosonic
Hubbard model was studied previously by a number of approximate methods. In
Ref.\ \cite{Singh} the real-space renormalization-group method was used,
with the result $(t/U)_c=0.215$ for the critical point. In Ref.\ \cite
{Krauth} Bethe-ansatz approximation for the full model was considered, with
a conjecture that the approximation actually corresponds to the exact
solution of the truncated model. The critical value obtained was $%
(t/U)_c=0.289$. In Ref.\ \cite{Freericks} the strong-coupling expansion for
the full model was examined. The results [$(t/U)_c=0.215$ for the bare
third-order expansion, and $(t/U)_c=0.265$ with a correction to the
Kosterlitz-Thouless critical behavior] are quite applicable to the truncated
model, since the triply and more occupied sites contribute only to the
higher-order terms.

The above-cited results are more or less close to each other, and to the
critical value of the {\it full} model, $(t/U)_c=0.304\pm 0.002$, obtained
with a controllable accuracy by an accurate method ``exact diagonalization +
renormalization-group (RG) analysis" \cite{Kashurnikov}. [Practically the
same value, $(t/U)_c=0.30$, was found also by density-matrix real-space RG
method \cite{Pai}, and by accurate MC simulations \cite{Krasavin}.] However,
the method of Ref.\ \cite{Kashurnikov} demonstrated that in the {\it %
truncated} model the critical value should be essentially shifted towards
small $U$'s: at $t/U=0.3$ the truncated system was found well inside the
insulating region. Moreover, the very fact of the existence of superfluidity
at $U>0$ was questioned [with a reservation that the maximal available size
of the cluster, $16$ sites, was not enough to resolve unambiguously the
behavior at small $U$'s.]

To clarify this situation, we perform an accurate MC study of the
commensurate truncated model. We start with the standard worldline MC method 
\cite{Batrouni} and the standard calculation of the insulating gap $\Delta
=\mu _{+}-\mu _{-}$, where the chemical potentials $\mu _{+}$ and $\mu _{-}$
are given by the relations $\mu _{+}=E(N_a+1)-E(N_a)$ and $\mu
_{-}=E(N_a)-E(N_a-1)$, $E(N_b)$ being the energy of a system consisting of $%
N_b$ bosons. In $\mu _{\pm }$ vs $t/U$ diagram the region of the insulator
phase is bounded by the curves $\mu _{+}(t/U)$ and $\mu _{-}(t/U)$. This
diagram for $N_a=N_b=50$ is shown in Fig.\ 1, in comparison with that for
the full system. At large $U/t$ phase boundaries in the truncated case are
close to those of the full one. This is the region, where the
strong-coupling approximation is rather accurate \cite{Freericks}. At
smaller $U/t$, however, the difference between the full and truncated models
is clearly seen. At $U/t<3.3$, where the full model is already in the
superfluid phase, the truncated model demonstrates a small, but quite
pronounced insulating gap. It is seen also, that the system is rather subtle
in this region, as the gap is considerably smaller than the hopping
amplitude. This feature makes it difficult to extract the critical point
from the data presented, even with the RG-analysis of the gap, since the
computational errors ($\sim 0.05t$) are comparable with the typical values
of the gap ($\sim 0.1t$) in a rather extensive critical region.

Fortunately, for the problem of pinpointing the critical ratio $U/t$ we had
an opportunity to take advantage of the continuous-time worldline MC
approach with ``worm'' update \cite{PST2}. The code described in Ref.\ \cite
{PST2} directly applies to our problem. Working in the grand canonical
ensemble, the code allows to extract the macroscopic parameter $K$ (main
superfluid characteristic of one-dimensional system \cite{Haldane}) from the
histograms for winding numbers and numbers of particles \cite{PST2}. Given
the values of $K$ for two different system sizes, one easily obtains
macroscopic value of $K$ from the first integral of the RG equations, and is
able thus to determine whether the system is superfluid or dielectric in the
macroscopic limit \cite{Kashurnikov}. The sizes of the systems we used for
this purpose were $N_a=128$ and $N_a=16$. This way we obtained the critical
point of the SF--MI transition in the truncated model. (The details of
calculation and the estimation of the error see in Ref.\ \cite{Kashurnikov}%
.) 
\begin{equation}
(U/t)_c=0.50\pm 0.05  \label{crit}
\end{equation}
As expected, this critical ratio is several times lower than that of the
full model.


\section{Attractive Regime}

\label{sec:3} Consider the truncated model with the attractive interaction, $%
U<0$. The behavior of the truncated model in this region differs
qualitatively from that of the full one, the latter being just trivial
because of the collapse of almost all of the particles onto one site. The
constraint on the occupation numbers renders the truncated model stable
against the collapse, so the point $U=0$ is not a special one for it.

In Fig.\ 2 we show the $\mu $ vs $U/t$ diagram for the truncated model in
the wide range of variation of $U/t$, including the attractive region up to $%
\mid U\mid /t=8$. In the region $-6.0<U/t<0.5$ the one-particle gap
vanishes, indicating that the ground state is an ordinary one-particle
superfluid. At $U/t<-6.0$ the one-particle gap ($\Delta \sim \mid U\mid $)
appears again. Apparently, the point $U/t<-6.0$ corresponds to the pairing
of the carriers. To demonstrate the liquid behavior of the two-particle
state, we calculate two-particle gap $\Delta _2$ [$\Delta _2=\mu _{2+}-\mu
_{2-}$, $\mu _{2+}=E(N_b+2)-E(N_b)$, $\mu _{2-}=E(N_b)-E(N_b-2)$]. The
results are presented in Fig.\ 3. These reveal the gap corresponding to
one-particle Mott insulator at $U/t>0.5$, and absence of the gap at negative 
$U$'s.

In Fig.\ 4b we show the one-particle gap at the critical region. Within the
accuracy of our results, the gap varies linearly with $U/t$.

Obviously, the pairing transition, being of liquid-liquid type, should take
place for non-commensurate fillings as well. [We checked this explicitly for
certain non-commensurate fillings.] Though not relevant for the pairing
transition, commensurability may lead however to the formation of the Mott
insulator for pairs. The absence of the two-particle gap in Fig.\ 3 at $U<0$
means that the state of two-particle Mott insulator does not take place,
except probably the case of very large $\mid U \mid$.

The limit of large attraction can be studied analytically. By the standard
second-order perturbative treatment in the parameter $t/U$ \cite
{Efetov,Emery} one obtains the following effective Hamiltonian in the
subspace of states with doubly occupied or empty sites: 
\begin{equation}
H_{\mbox{\scriptsize eff}}=\frac{t^2}{\mid U\mid }\sum_{i=1}^{N_a}\left\{
-(b_i^{\dag }b_{i+1}+\mbox{H.c.})+2b_i^{\dag }b_ib_{i+1}^{\dag
}b_{i+1}\right\} \;,  \label{H_eff}
\end{equation}
where $b_i^{\dag }$ creates a pair on the site $i$. The states with single
occupations of sites are separated from the ground state by the gap close to 
$\mid U\mid $. We thus come to the familiar 1D half-filled hard-core model
with the parameters corresponding just to the critical point of superfluid
-- insulator (commensurate density wave) transition (see, e.g., \cite
{Scalettar}). So we conclude that in our model the ``critical point'' for
the Mott transition for pairs is $U=\infty $, and for finite $U$'s the
transition does not take place.


\section{Response to Gauge Phase}

\label{sec:4} Pairing leads to a dramatic change in the response of the
system to the global gauge phase. It is instructive thus to study this
response in the vicinity of the critical point.

We introduce the gauge phase, $\Phi$, to the model by conventional
transformation of the hopping amplitude: $t \rightarrow t \exp (i \Phi /N_a)$%
.

First we study the groundstate energy, $E_0$, as a function of $\Phi$, at
different $U$'s, by exact diagonalization for the chain $N_a=N_b=12$, see
Fig.\ 5. [The results for the groundstate persistent current, $J_0(\Phi)$,
immediately follow from $J_0(\Phi) = d E_0(\Phi) / d\Phi$.] Curve 1
corresponds to the well-defined one-particle ground state laying in the zero
momentum sector at any value of $\Phi$. The well-defined two-particle ground
state (curve 4) lays in the zero sector while $\Phi$ varies from $0$ to $\pi
/2$. At $\Phi=\pi /2$ it jumps to the momentum sector $N_b/2$ (in the units
of minimal non-zero momentum), and remains there till $\Phi=3\pi /2$. The
curve $E_0(\Phi)$ acquires a characteristic cusp form, symmetric with
respect to the line $\Phi =\pi /2$. Curves 2 and 3, corresponding to the
cross-over region, demonstrate a marginal behavior of $E_0(\Phi)$: The
ground state lays in the zero momentum sector until the phase reaches some
critical value (greater than $\pi /2$), and then jumps to the sector $N_b/2$%
. The curve is not symmetric with respect to the cusp point.

We see that the marginal response to the gauge phase in the vicinity of the
critical point is due to the competition (at $\Phi > \pi /2$) between the
groundstate in zero sector and that in the sector $N_b/2$. At finite, but
sufficiently low temperatures both states contribute to the marginal
response (provided the system is close enough to the critical point), with
their Gibbs factors. As an example, we present in Fig.\ 6 the equilibrium
persistent current as a function of $\Phi$ obtained by MC.

The marginal response to the gauge phase can be used to obtain a rather
accurate extrapolation for the critical point. At $N_a \rightarrow \infty$
the size of the cross-over region, where the marginal response takes place,
tends to zero. Hence, we may consider the value of $U$, corresponding to
some characteristic marginal-response feature, as a function of $N_a$, and
extrapolate to $N_a = \infty$. For one thing, one may consider just the
point of the appearance of the marginal response, i.e. the point where at
the phase $\Phi=\pi$ the energies in the zero sector and in the sector $%
N_b/2 $ coincide. Corresponding extrapolation from the exact-diagonalization
results is presented in Fig.\ 7. The difference between the results of
linear and quadratic extrapolations provides a reasonable estimate for the
error.


\section{Spin-1 Chain with Axial Anisotropy}

\label{sec:5}

Macroscopically, the truncated bosonic model considered (TBM) in the
previous sections is analogous to the spin-1 chain with axial anisotropy
(S1): 
\begin{equation}
H=-\frac t2\sum_{\left\langle ij\right\rangle }\left(
S_i^{+}S_j^{-}+S_i^{-}S_j^{+}\right) +V\sum_{\left\langle ij\right\rangle
}S_i^zS_j^z+U\sum_i\left( S_i^z\right) ^2\,\,,  \label{antiferr}
\end{equation}
where $U$ is the parameter of the interaction with the easy-magnetization
axis, if $U<0$, or easy-magnetization plane, if $U>0$. [Generally speaking,
the term with $V$ renders the model (\ref{antiferr}) much reacher than the
model (\ref{Hub}). Obviously, to make the analogy more close, one should
also add a term with nearest-neighbor interaction to the Hamiltonian (\ref
{Hub}). To our purposes, however, this is not essential].

The zero-point phase diagram of the model (\ref{antiferr}) was studied in
Refs.\ \cite{Botet,Solyom} by exact diagonalization on the chains up to $%
N_a=12$, see Fig.\ 8. The phases were identified as follows:

1) ferromagnetic ($\langle S_z \rangle =\pm N_a$);

2) antiferromagnetic ($\langle S_z \rangle =0$ in the ground state as well
as in the first excited one);

3) Haldane gap state (the state with the gap in the spectrum, and $\langle
S_z \rangle =\pm 1$ in the first excited state);

4) $XY$-spin liquid (the gapless state with $\langle S_z \rangle =\pm 1$ in
the first excited state).

[Haldane gap state in S1 is equivalent to Mott state in the TBM, while and
the $XY$-spin liquid in S1 is equivalent to one-particle superfluid in the
TBM.]

Besides, in Ref.\ \cite{Botet} one more phase was observed, the so-called
``spin-$1/2$-like $XY$-phase'' (region 5 in Fig.\ 8), which was
characterized by $\langle S_z \rangle =0$ in the ground state and $\langle
S_z \rangle =\pm 2$ in the first excited state. But later, in Ref.\ \cite
{Solyom} this phase was not detected, and, as far as we know, it had not
been mentioned in literature since that time.

Our aim is to make sure that this phase does really exist and is equivalent
to the two-particle superfluid in TBM. To this end we perform MC simulation
of the model (\ref{antiferr}) in the bosonic representation
(Holstein-Primakoff transformation) for $N_a=50$, at $V=-0.05t$. The $\mu $
vs $U/t$ diagram is shown in Fig.\ 9. It demonstrates the remarkable
similarity with the behavior of TBM near the point of the pairing phase
transition: the one-particle gap, which shrinks linearly in the wide
interval of the parameter $U$. The critical point $(U/t)_c\approx -2.0$
coincides with the value found in Ref.\ \cite{Botet}.

Note that in the limit $U \rightarrow \infty$ the model (\ref{antiferr})
with $V=0$ also approaches the effective Hamiltonian (\ref{H_eff}), and the
equivalence to TBM becomes exact.


\section{Acknowledgements}

\label{sec:6}

The authors are grateful to N.\ Prokof'ev for useful discussions. This work
was partially supported by the Russian Foundation for Basic Research [Grants
No. 95-02-06191a and No. 97-02-16187]. B.V.S. acknowledges also the support
by Grant No. INTAS-93-2834-ext [of the European Community].

\newpage\ 

FIGURE CAPTIONS

\mbox{}\\Fig. 1. $\mu $ vs $U/t$ diagram for the truncated model ($\circ $)
with $N_a=N_b=50$ obtained by the Monte Carlo method. $T=0.0625t$. For
comparison the phase diagram of the full boson model ($+$) is shown \cite
{Krasavin}.

\mbox{}\\Fig. 2. $\mu$ vs $U/t$ diagram, obtained for $N_a=N_b=50$ by Monte
Carlo in the wide parameter region.

\mbox{}\\Fig. 3. $\mu_2$ vs $U/t$ diagram for $N_a=N_b=14$ (exact
diagonalization).

\mbox{}\\ Fig. 4. One-particle gap at the critical region $T=t/50$, $N_a=50$
(Monte Carlo).

\mbox{}\\ Fig. 5. Groundstate energy vs gauge phase at $N_a=N_b=12$ (exact
diagonalization): (1) $U/t=-5.5$, (2) $U/t=-6.0$, (3) $U/t=-6.5$, (4) $%
U/t=-10.0$.

\mbox{}\\ Fig. 6. Equilibrium persistent current vs gauge phase in the
marginal region. $U/t=-5.92$, $N_a=50$, $T=t/150$.

\mbox{}\\ Fig. 7. Scaling for the point of the appearance of the marginal
response to the gauge phase for chains with $N_a=8,\,10,\,12,\,14,\,16$
(exact diagonalization).

\mbox{}\\Fig. 8. Phase diagram for spin-1 chain with axial anisotropy, from
Ref.\ \cite{Botet,Solyom}. (1) ferromagnetic state, (2) antiferromagnetic
state, (3) Haldane gap state, (4) $XY$-spin liquid, (5) ``spin-$1/2$-like $%
XY $-phase''.

\mbox{}\\ Fig. 9. $\mu $ vs $U/t$ diagram for the spin chain. $V/t=-0.05$, $%
N_a=50$, $T=t/50$.

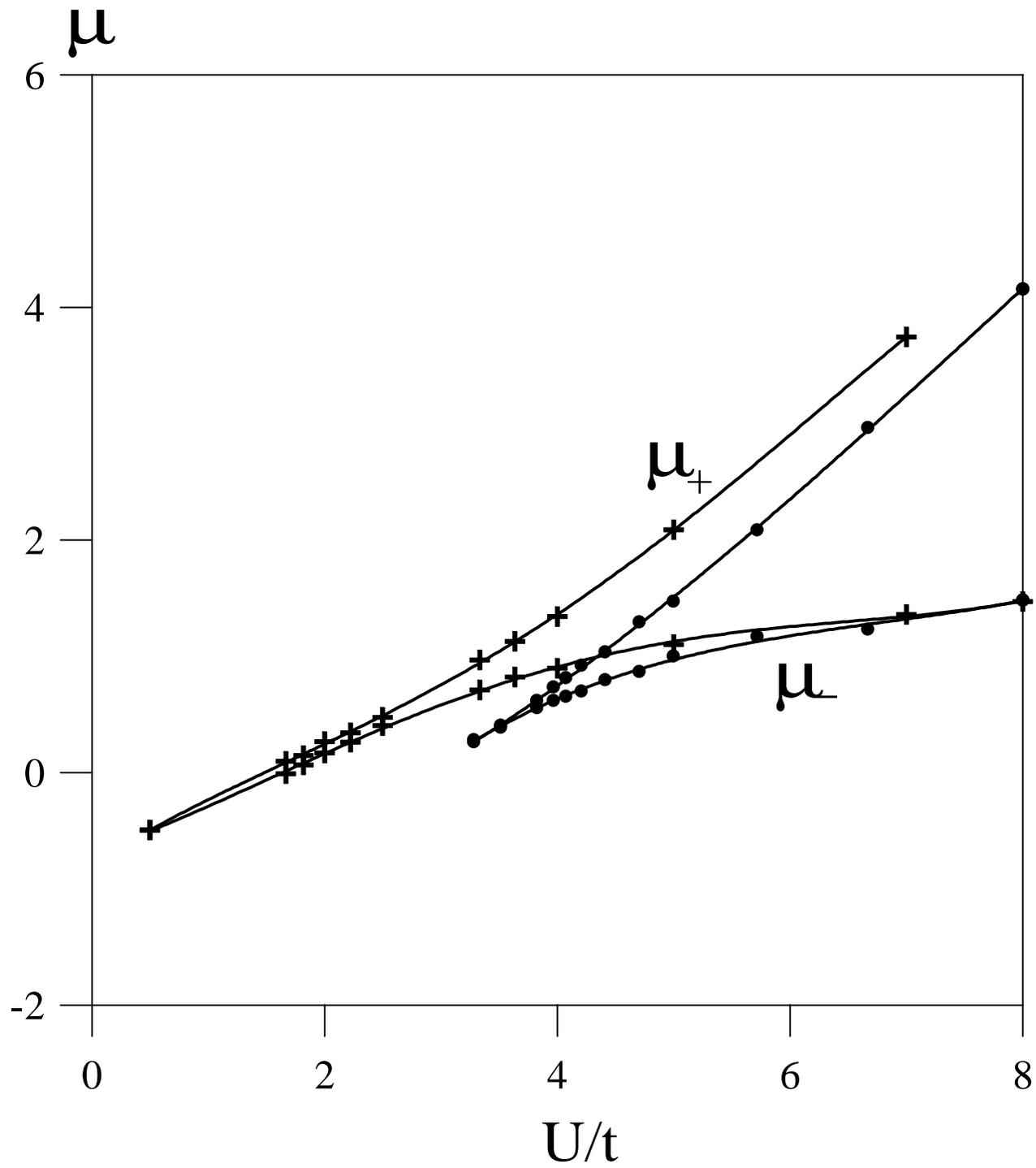
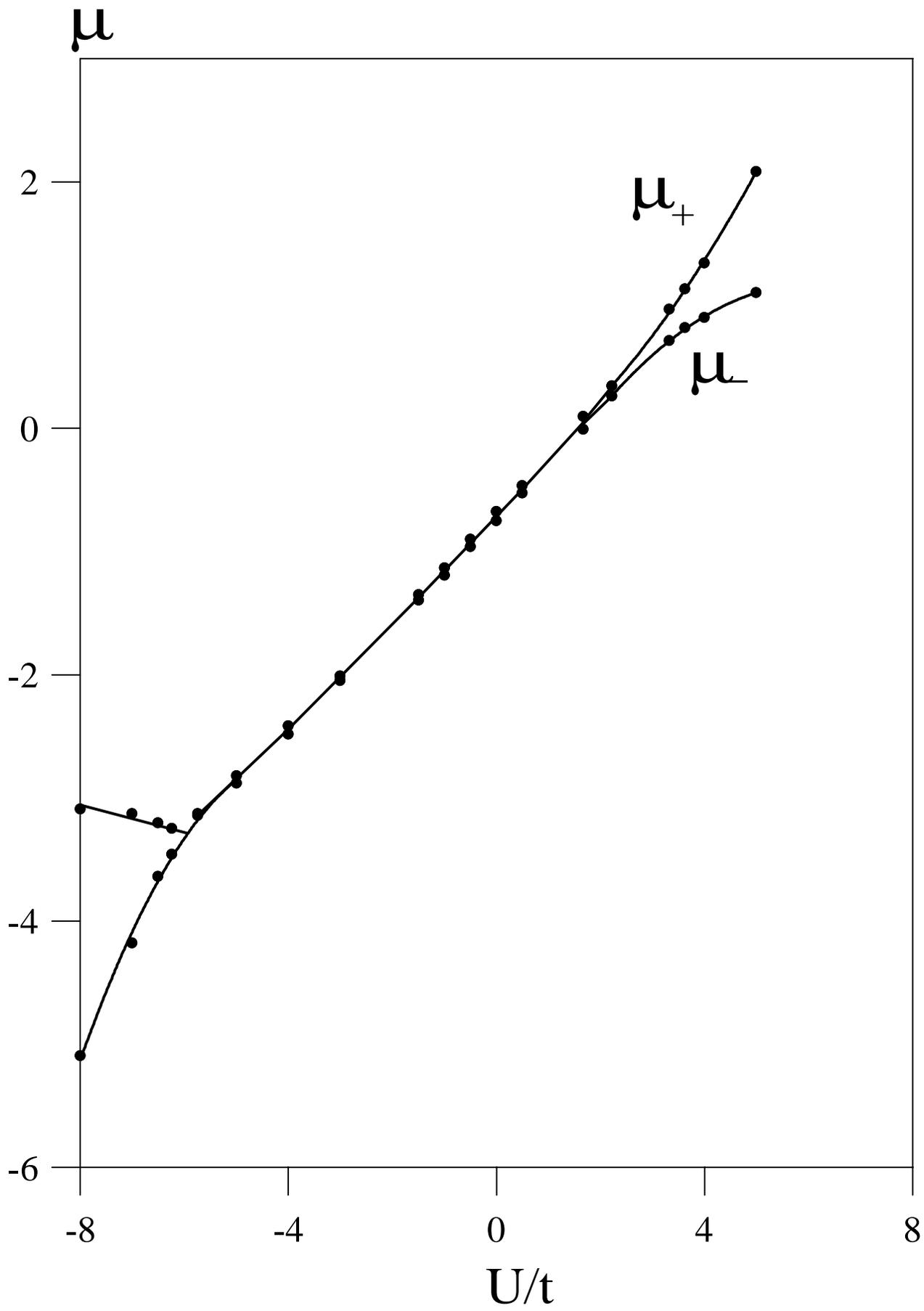
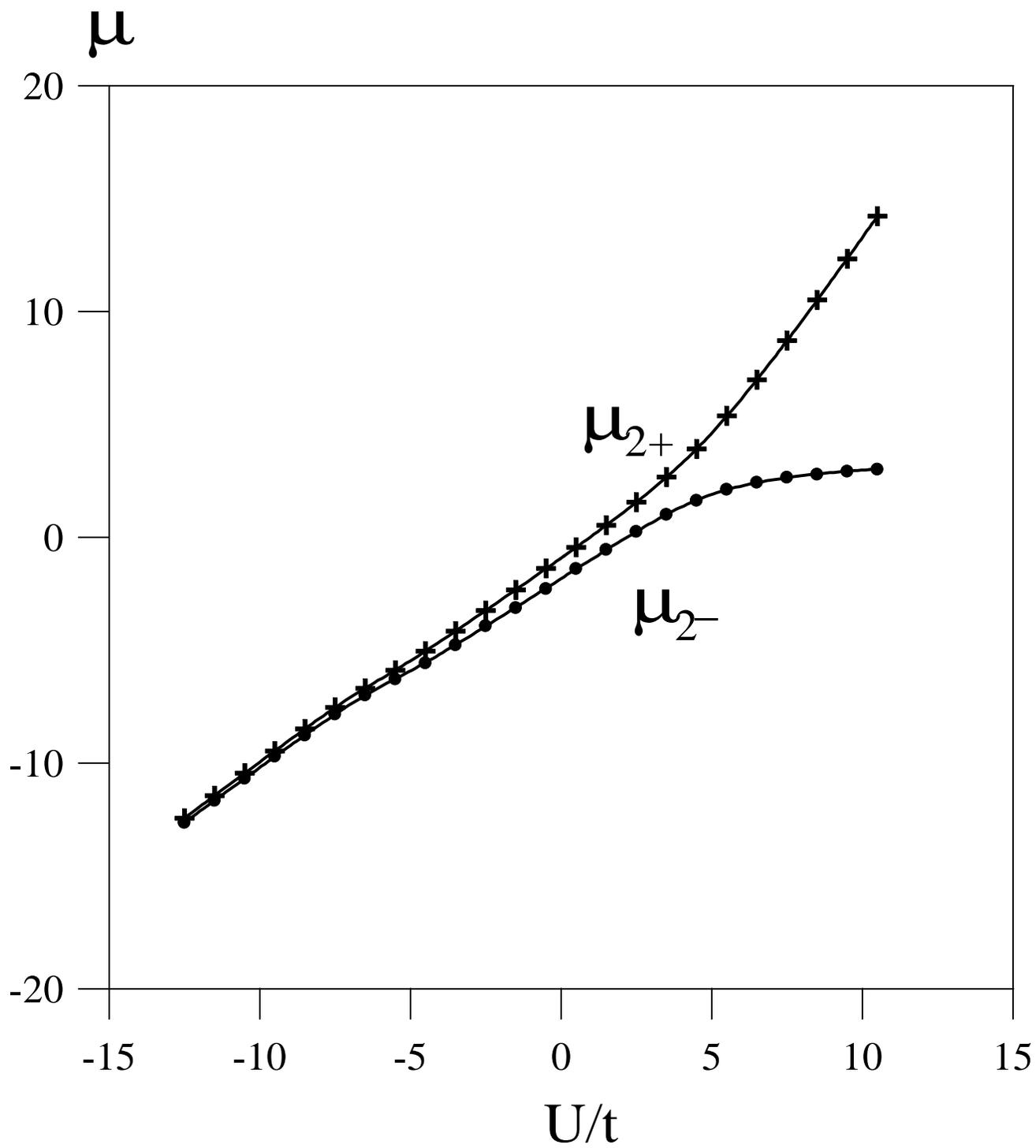
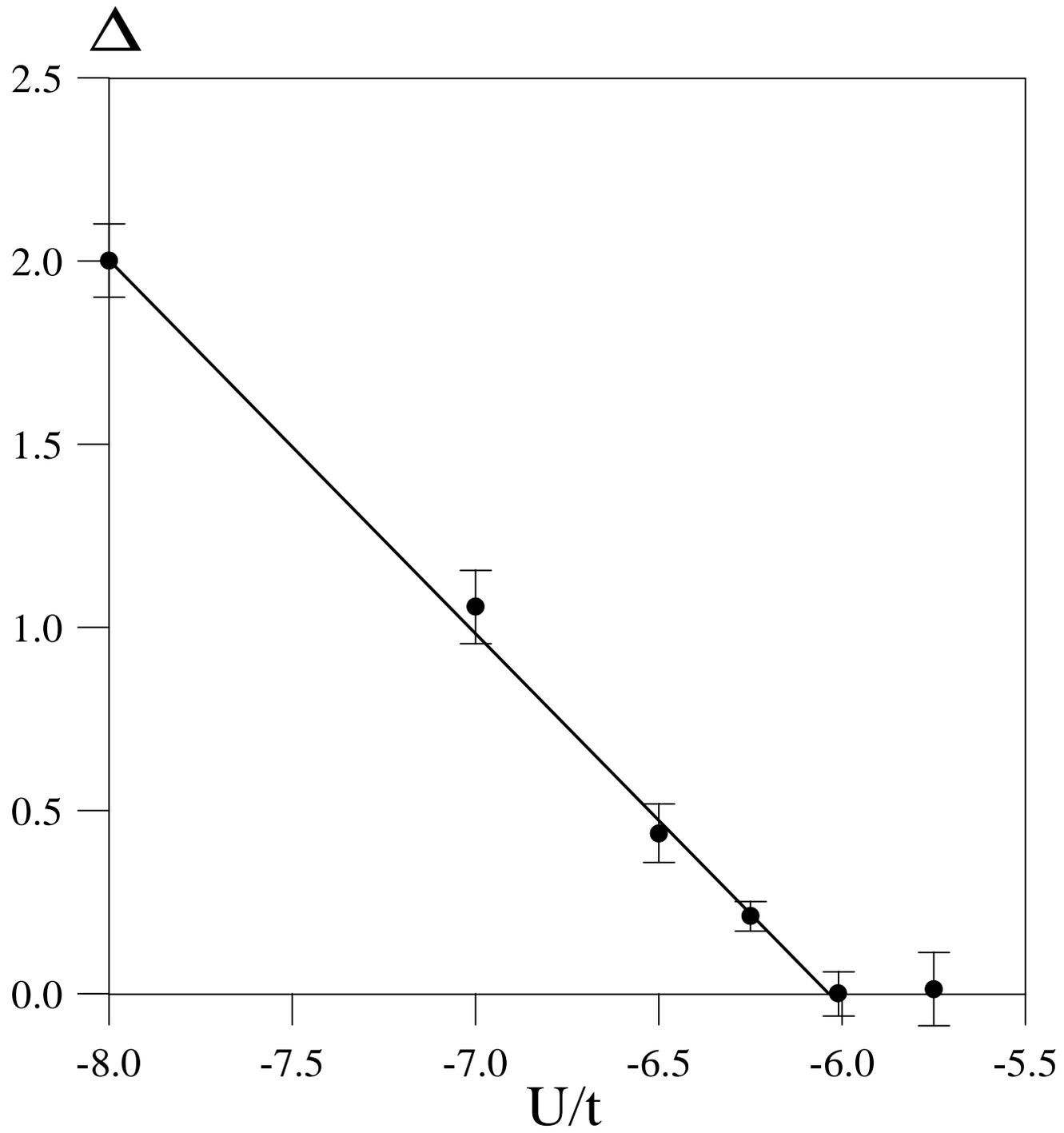
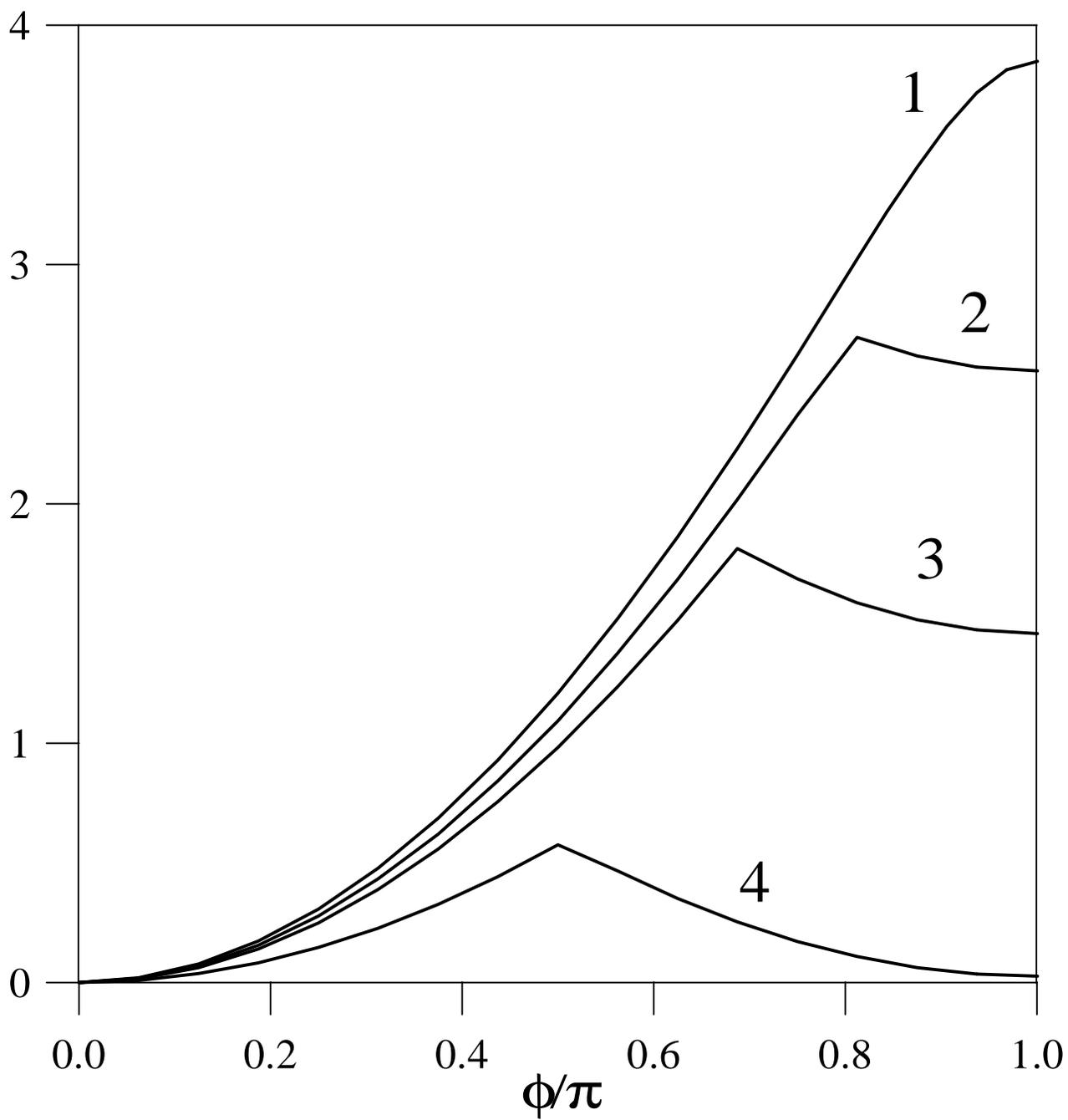
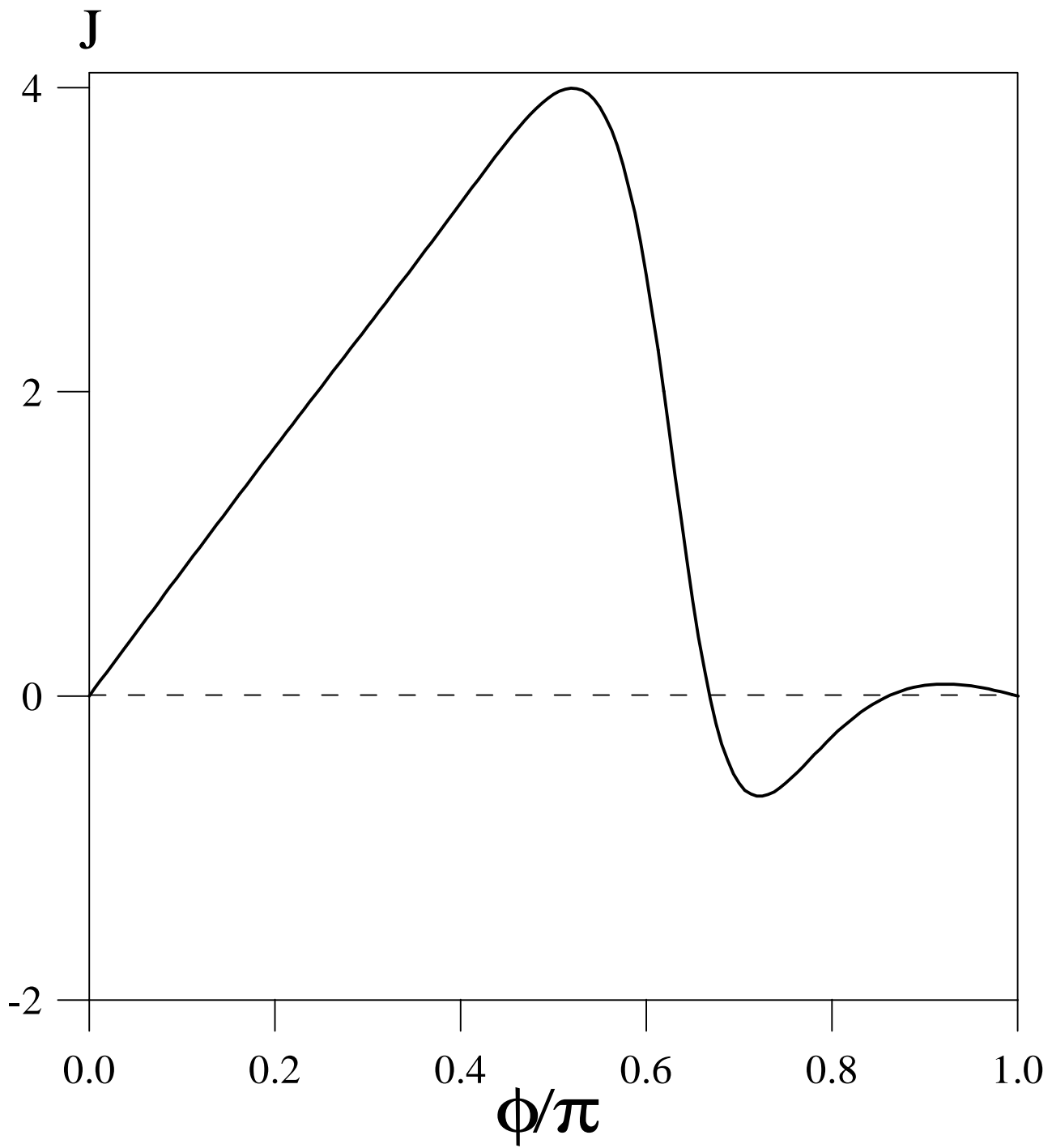
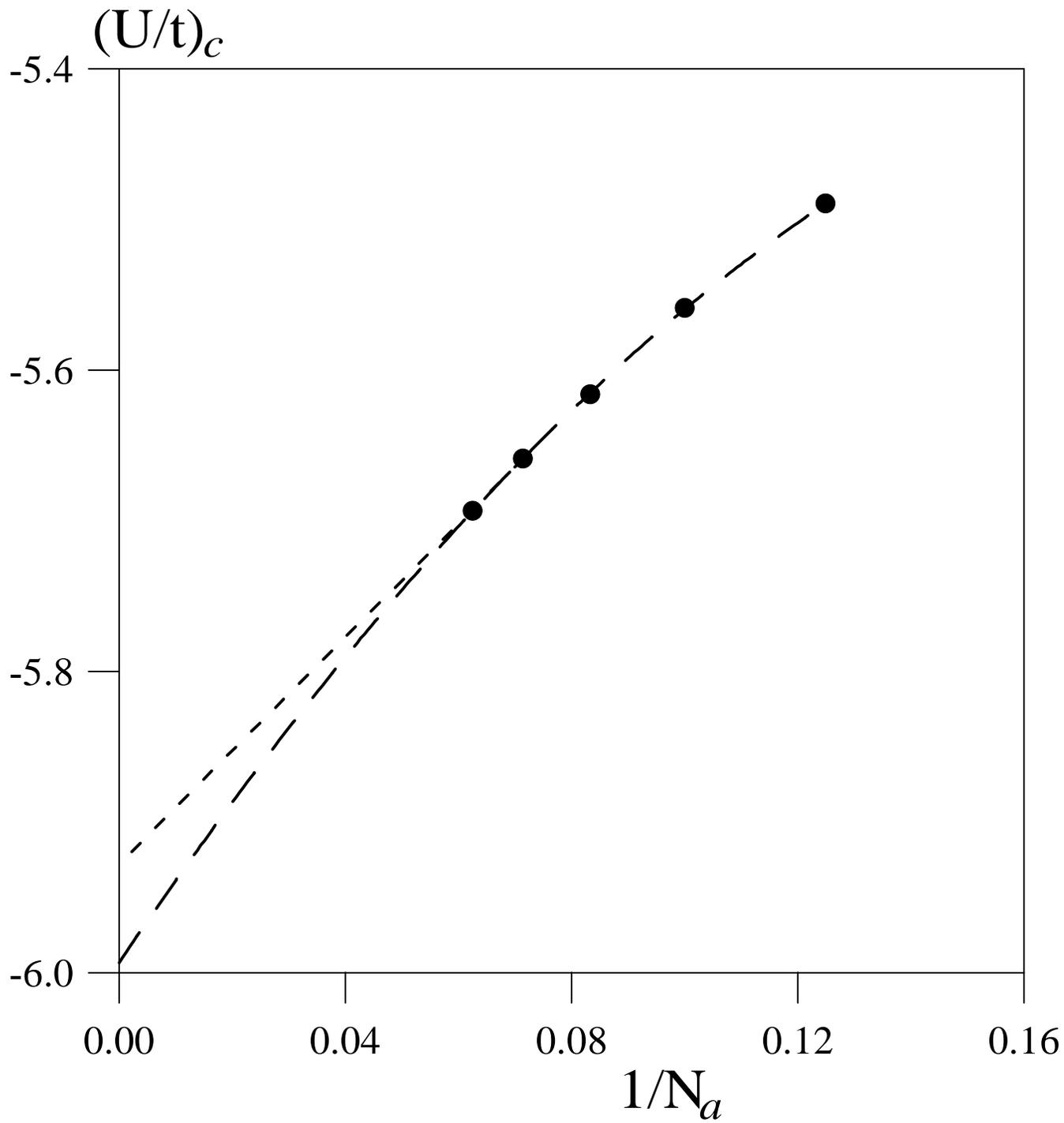
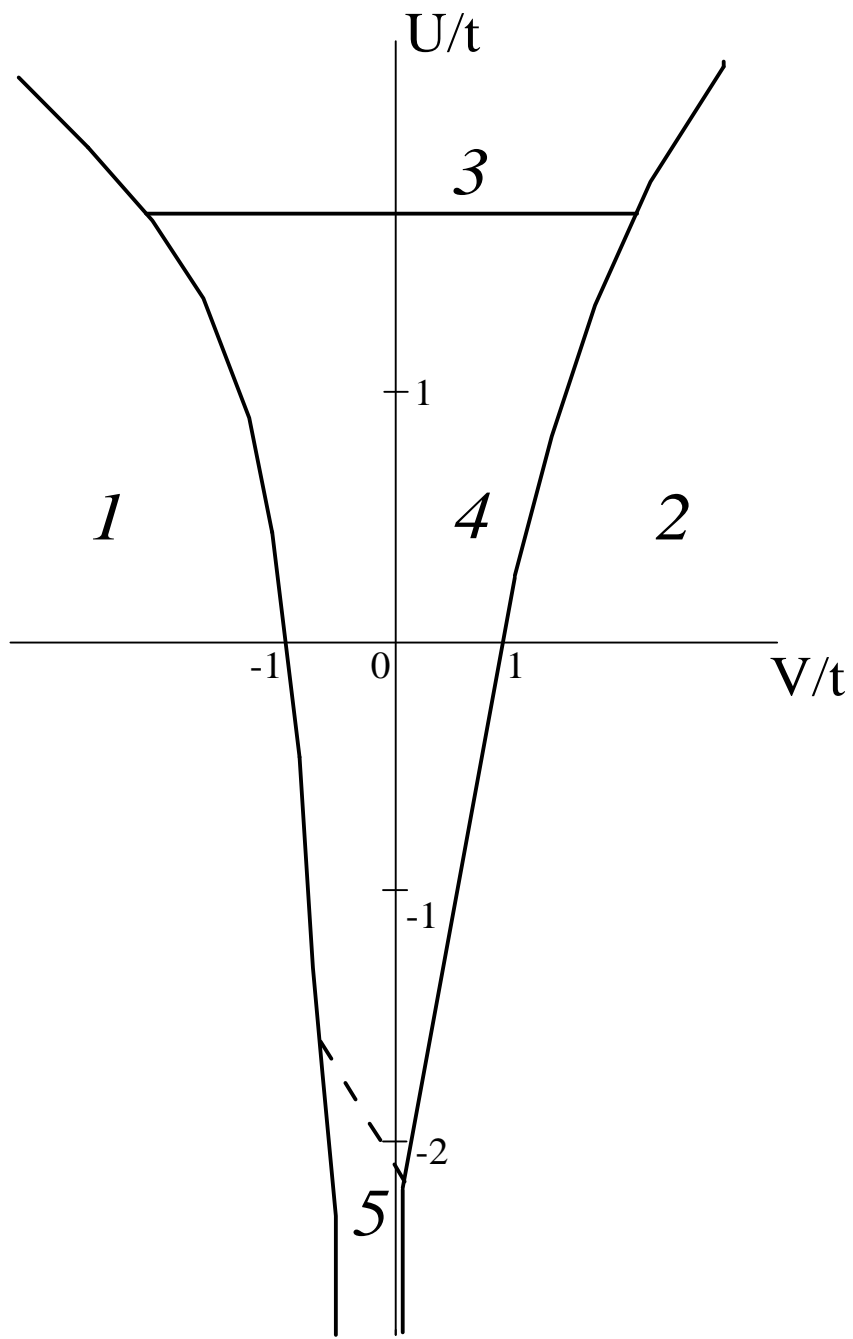
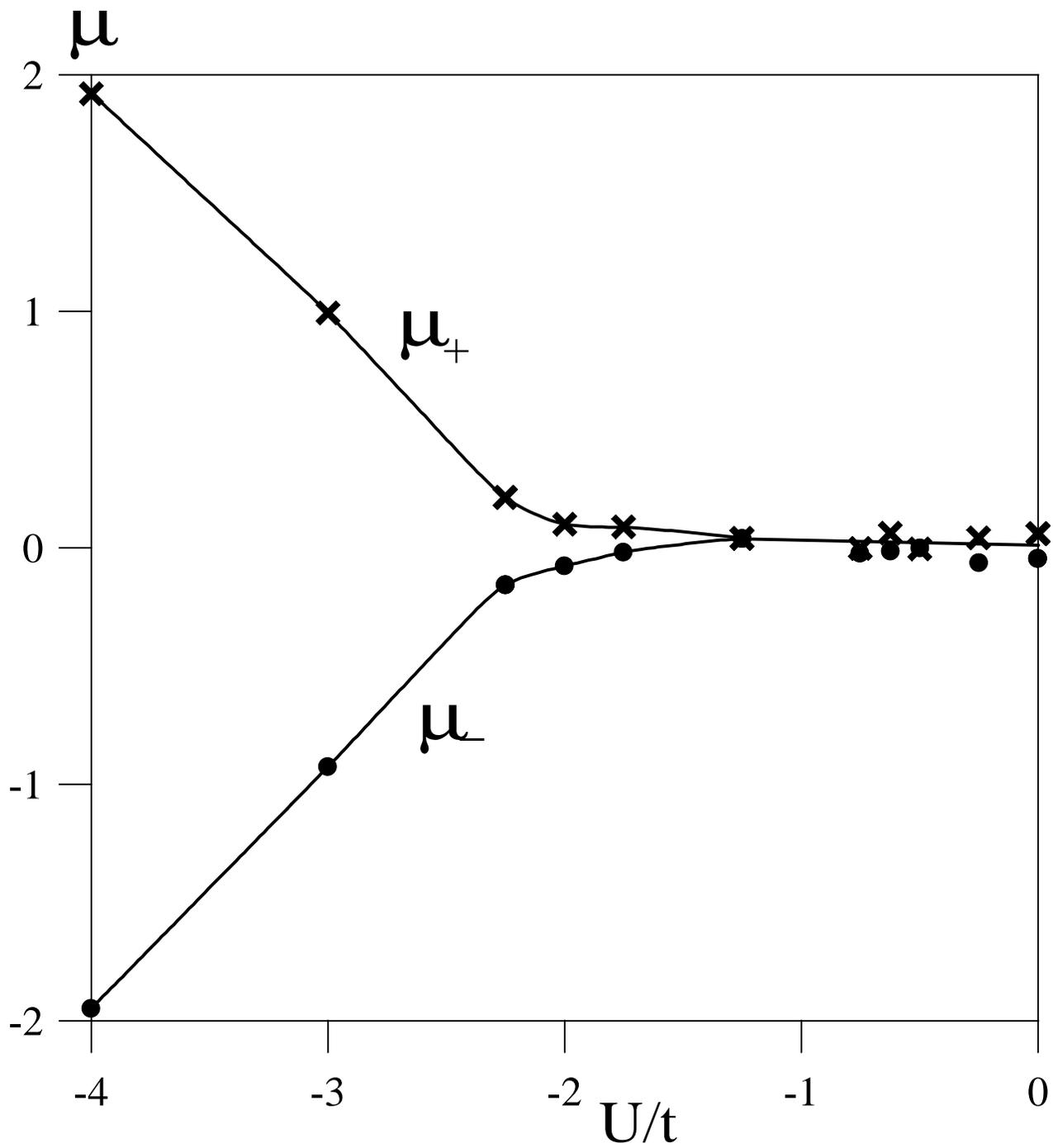
\end{document}